\documentclass[12pt]{article}

\input epsf

\begin{document}
\baselineskip 14pt
\title{The Reality in Bohmian Quantum Mechanics or Can You Kill with an Empty Wave Bullet?}

\author{ Lev Vaidman}

\maketitle
\vspace{.4cm}
\centerline{ School of Physics and Astronomy} 
\centerline{Raymond and Beverly Sackler Faculty of Exact Sciences}
\centerline{Tel-Aviv University, Tel-Aviv 69978, Israel}

\date{}

\vspace{.2cm}
\begin{abstract}
  Several situations, in which an empty wave causes an observable
  effect, are reviewed. They include an experiment showing
  ``surrealistic trajectories'' proposed by Englert et al. and
  protective measurement of the density of the quantum state.
  Conditions for observable effects due to empty waves are derived.
  The possibility (in spite of the existence of these examples) of
  minimalistic interpretation of Bohmian Quantum Mechanics in which
  only Bohmian positions supervene on our experience is discussed.

\end{abstract}
\vspace {.1cm}


\vskip 1cm

\break
\section{INTRODUCTION}
\label{intr}

I have my most vivid memories of Jim Cushing from ``Bohmian
conference'' in 1995 which took place in Bielefeld, Germany and had
the title ``Quantum Theory without Observers''. The majority of
participants were devoted Bohmians and most of the talks and
discussions were about meaning and achievements of Bohmian quantum
mechanics \cite{Bohm,Gold}. At that time I already was an enthusiastic proponent of the
Many-Worlds interpretation (MWI) \cite{Eve}, and Bohmian interpretation, which  close to the MWI, was very intriguing for me. Illuminating
discussions with Jim Cushing led me to think more about Bohmian
Interpretation.

I am still a strong proponent of the MWI \cite{MWIStan}. The main reason for this is not
the philosophical advantage of the plurality of worlds, but a desire to view
physics as a theory of everything. The main obstacle for this is the
collapse of the quantum wave. Collapse introduces randomness into
physics, it puts limits on predictive power of physics. There is no attractive
proposal for a physical theory of collapse, and, moreover, it seems impossible to
define when collapse occur. I get used to the idea of plurality of
worlds, but a theory without collapse and with a single world is clearly a better
theory of everything. In some sense, Bohmian quantum mechanics is such 
theory. (Note, however, that Bohm himself never viewed his theory in
that way. I had elaborate discussion with him in South Carolina in 1989
in which he explained that his theory is another  step in the
evolution of physics and there will never be the final theory of
everything.) The main reason (apart from nonlocality of Bohmian
mechanics) why I still prefer the MWI, is that it does not really
eliminate the plurality of worlds. The formalism still has many, many
``empty wave'' worlds in which I walk, eat, slip and, in particular,
write papers.  Nevertheless, Bohmian mechanics achieves something that
no other theory was able to do: to single out in a pretty natural way
a single world out of the plurality of worlds in the MWI. (Note that in some special
cases the Bohmian world might slightly differ from the world of the MWI
\cite{AVBohm}).

In the MWI, the Wave Function of the Universe is decomposed into
superposition of branches in which  the shape  of the wave
function yield a sensible picture, and the time evolution of the wave
function of a branch yields a sensible story (with possible further
branching). It is  postulated that our experience corresponds to all
branches with sensible stories. 

In the Bohmian mechanics, or, at least, in my approach to Bohmian
mechanics, it is postulated that our experience corresponds to Bohmian
positions. Bohmian positions correspond to a sensible picture in three
dimensions which evolves in time and yields a sensible {\it single}
story. The same arguments which start from the locality of known interactions
(which are frequently named as decoherence theory) yield plurality of particular
sensible stories in the MWI and a single story (usually identical to one of
the MWI stories) in the Bohmian mechanics.

In my approach  our experience is related solely
to Bohmian positions and not to the quantum state (the wave
function). It is not a new  approach: it seems to me
that it is the {\it pilot wave} approach as Bell \cite{Bell} understood it, and maybe
De Broglie imagined it. Bedard \cite{Beda} attributes this view also.
to Holland \cite{Ho}, Maudlin \cite{Mad} and Albert \cite{Alb}.

In this paper I am going to analyze the significance of empty waves in
the light of recent results about position measurements which do not
show Bohmian positions. I will conclude with discussion of the
interpretation, advocating minimalistic approach according to which
our experience supervene only on Bohmian positions (and not on the
quantum wave). I add a short discussion of Bedard's arguments against
minimalistic interpretation in the appendix.

\vskip 1cm

\section{Can an empty wave kill?}

Consider a gedanken experiment in which a bullet splits its quantum
wave at a ``beam splitter'' into two equal weight wave packets, one
moving toward a cat, while another misses the cat. For simplicity, we consider
equal, uniform density, spherical wave packets.
Consider a situation that the Bohmian position
of the bullet is inside the wave packet that misses the cat, see Fig. 1.
It was generally believed that in this situation we should not 
worry about cat's health. However, recently, there were several works showing that
empty waves can have observable effects. Hardy \cite{Ha92} discussed
empty waves in the interaction free measurements \cite{EV93}, Englert
et al. \cite{Sur} discovered ``Surrealistic Bohmian trajectories'' in
which an empty wave leaves a trace of its trajectory and Aharonov et al. \cite{PMBohm1,PMBohm2} showed that in protective measurements \cite{PM,PM1} one can observe
the shape of the quantum wave, while the Bohmian particle is
essentially at rest and does not visit the regions where the value of
the wave function is measured.

\begin{center} \leavevmode \epsfbox{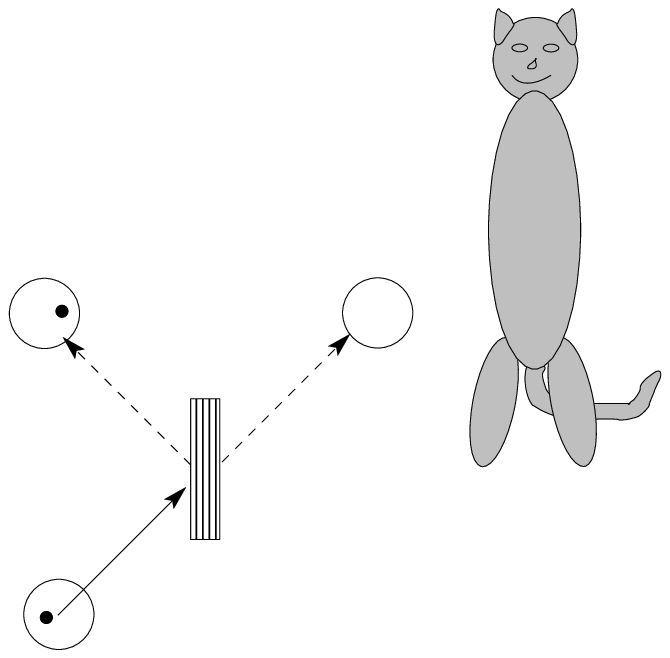} \end{center}

\noindent 
{\small Fig. 1. A bullet is fired toward a cat, but after the
  beamsplitter only an  empty wave of a bullet comes toward him. Bohmian particle position, signified by a black dot travels away with the other part of the quantum wave. }

\vskip .5cm

 An empty wave can,
in the future, reach the Bohmian position of the particle and cease to
be ``empty''. Clearly, at this stage it can lead to observable
changes, i.e. to change Bohmian trajectory. Bell \cite{Bell} pointed
out that usually it will be a dominant influence. Consider the two  wave packets of the ``bullet'' which are forced to overlap
again (as in a Mach-Zehnder interferometer without second
beam-splitter), see Fig. 2. At the point of the overlap, the empty
wave ``grabs'' the Bohmian particle. Indeed, this result can be
immediately seen from Bell's presentation of Bohm's theory in the form
of the pilot wave where the velocity of Bohmian position of a particle
depends on the current density and the wave density at the location of
the Bohmian particle:
\begin{equation}
\vec{v}={\vec{j} \over \rho} \ , \label{eqmot}
\end{equation}
where $\rho(\vec{x} ) =
|\psi(\vec{x} )|^2$, and $\vec{j}=\frac{\hbar}{2 i m}
\{\psi^*\vec{\nabla}\psi-\psi\vec{\nabla}\psi^*\}$. 
The velocity in the region of the overlap of the two wave packets is
given by 
\begin{equation}
\vec{v}={{\vec{j_1}+\vec{j_2}}\over {  \rho_1 + \rho_2}}={{\vec{v_1} + \vec{v_2}}\over 2} \ , \label{eqmot12}
\end{equation}
where indexes ``1'' and ``2'' correspond to the two wave packets. The
horizontal component of the velocity of the Bohmian particle vanishes
during  the time it is inside both wave packets. So, from the moment
of the overlap starts the competition between the two wave packet:
which one will keep the point inside it longer? When one wave
packet leaves, the point continue to move with the velocity of the
second wave packet and it remains inside it. Since at the beginning of
the overlap, the Bohmian position is at the boundary of the empty wave
packet and it is inside the other one, the empty wave has longer way to go and it always ``wins'' the
competition: the empty wave ``grabs'' the Bohmian particle.

\begin{center} \leavevmode \epsfbox{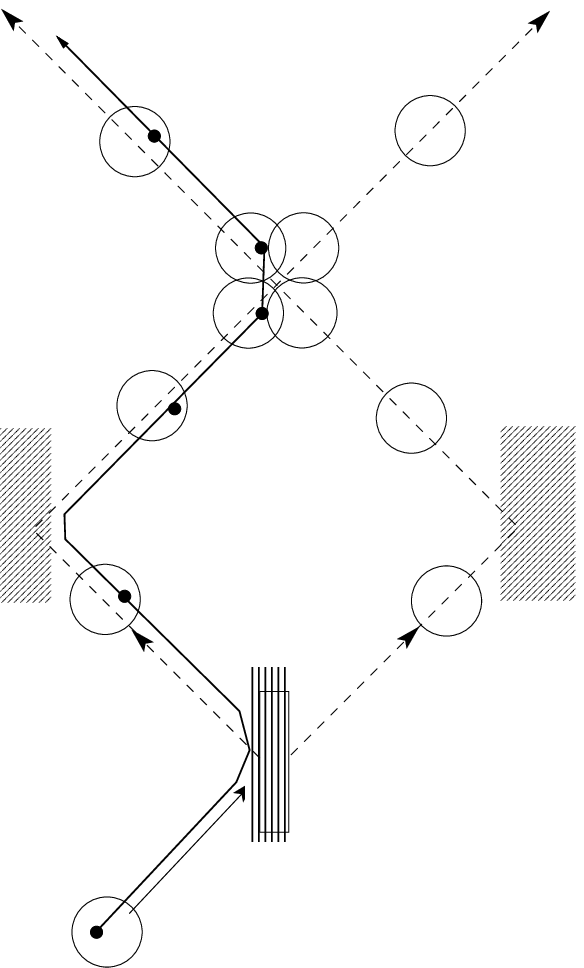} \end{center}

\noindent 
{\small Fig. 2. Trajectory of a Bohmian particle in a Mach-Zehnder
  interferometer without second beamsplitter. At the meeting point of
  empty and nonempty wave packets, the Bohmian particle ``changes
  hands'' and continues to move with what was before an empty wave.}

\vskip .5cm

The situation is  different if the empty wave bullet on its way
``kills'' an ``empty wave'' cat, see Fig. 3. Even if the physics is such that the
bullet goes through the cat without significant delay and the empty wave packet
of the bullet comes in time to overlap with the non-empty wave
packet, the empty wave does not grab the Bohmian particle in this
case. Indeed, the velocity of the
Bohmian  particle  in the region of the overlap is   
\begin{equation}
\vec{v}={{\vec{j_1}}\over {  \rho_1 }}=\vec{v_1},   \label{eqmot13}
\end{equation}
where $\vec{v_1}$ is the velocity of the non-empty wave packet.
The reason is that at the time of the overlap of the wave packets of
the bullet, the wave packets of some parts of the cat's body do not
overlap. The wave packets of these parts, entangled with an empty wave bullet, move relative to the case of the  undisturbed cat.  The Bohmian positions of the particles in the cat's body are that of an uninjured
cat, and therefore, the wave packets of some particles of the body entangled with the empty wave
packet of the bullet do not contain the Bohmian particle inside it. In
the configuration space of all involved particles (of the bullet and of
the cat) there are two wave packets but the Bohmian position belongs
only to one of them.  

\begin{center} \leavevmode \epsfbox{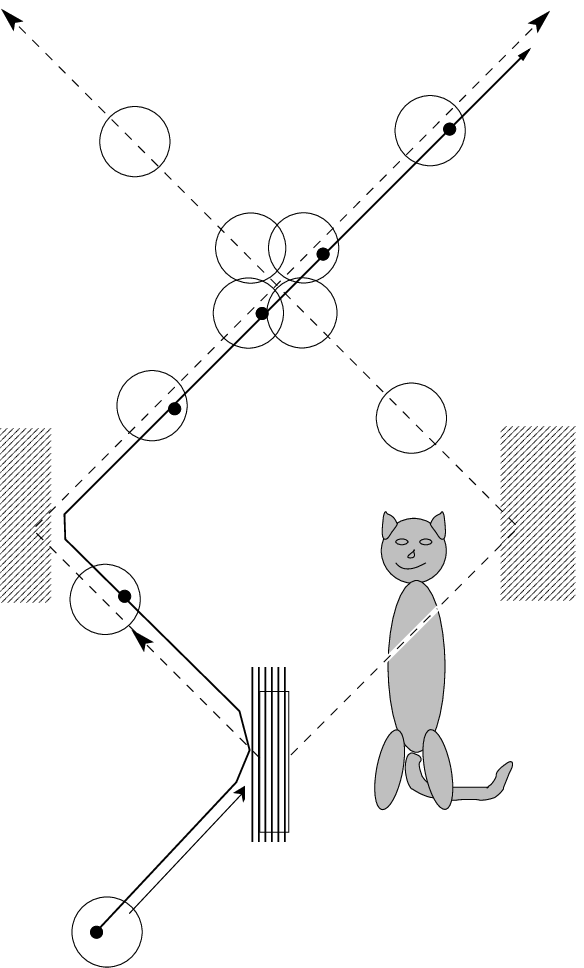} \end{center}

\noindent 
{\small Fig. 3. Trajectory of Bohmian particle in a modified
  experiment. When the empty wave ``kills'' on its way, it does not
  influences the Bohmian trajectory in the future even if it overlaps
  with the nonempty wave as it was in Fig. 2.}

\vskip 1cm

\section{Surrealistic trajectories}

The examples presented above are not too surprising (although we are
not used in Newtonian mechanics to change in velocity without
interaction as in Fig. 2): the empty wave influences other objects
only when it cease to be an empty wave. A really surprising result was
discovered by Englert et al. \cite{Sur}. They realized that if (instead
of killing a cat) the bullet will flip spins on its way, then the
Bohmian trajectory will be as if the wave packets move in a free
space, see Fig. 4. Nevertheless, the flipped spins show a different
trajectory. There were many  discussions regarding the meaning and
significance of this example \cite{ST1,ST2,ST3,ST4,Bar}. When I first heard about this
result, I did not believe it until I checked it myself.

\begin{center} \leavevmode \epsfbox{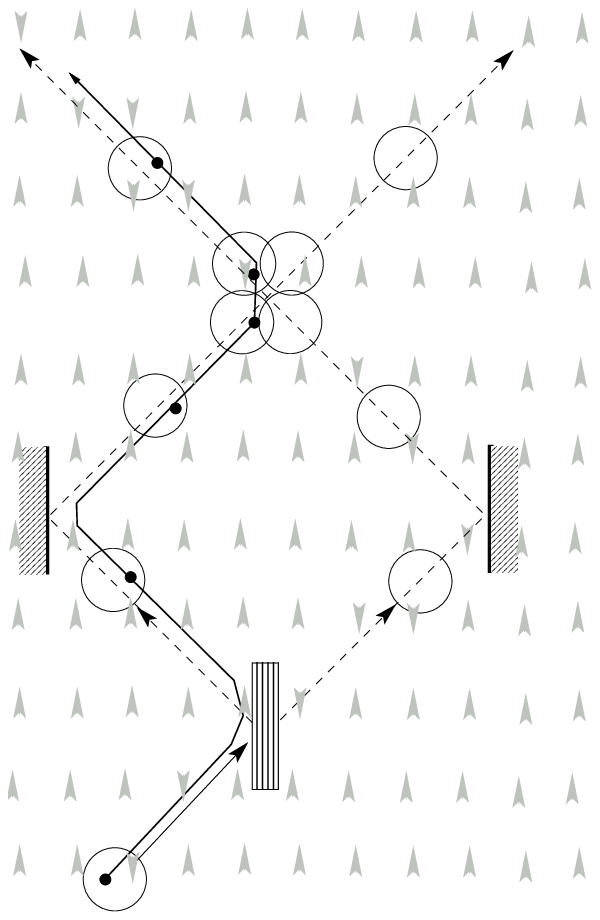} \end{center}

\noindent 
{\small Fig. 4. Surrealistic trajectories. The trace of flipped spins and the actual trajectory of Bohmian particle are different.  Which one is ``surrealistic'' is a matter of interpretation. (Note that flipped quantum waves of  spins are only in one branch of the universal wave function, the branch corresponding to particular detection of the particle in the detector on the left.}

\vskip .5cm

 My modification of this idea is
to consider a very fast particle moving in a special bubble chamber in
which the bubbles are developed slowly.
 During the time the particle moves inside the interferometer, the
 quantum states of electrons of excited atoms which later  create the
 bubbles have no enough time to move out of the Bohmian positions of
 the electrons that are at rest at this time. The electron Bohmian positions are at rest because the excited
 states do not contribute to  the Bohmian velocity when the Bohmian
 position of the wave packet of the particle moves in another place.
The result of the experiment (which can be seen only much later) is a
trace of bubbles corresponding to one trajectory while the trajectory
of the Bohmian
position  is the other one, Fig. 5. The bubbles show the trajectory
of the empty wave!

\begin{center} \leavevmode \epsfbox{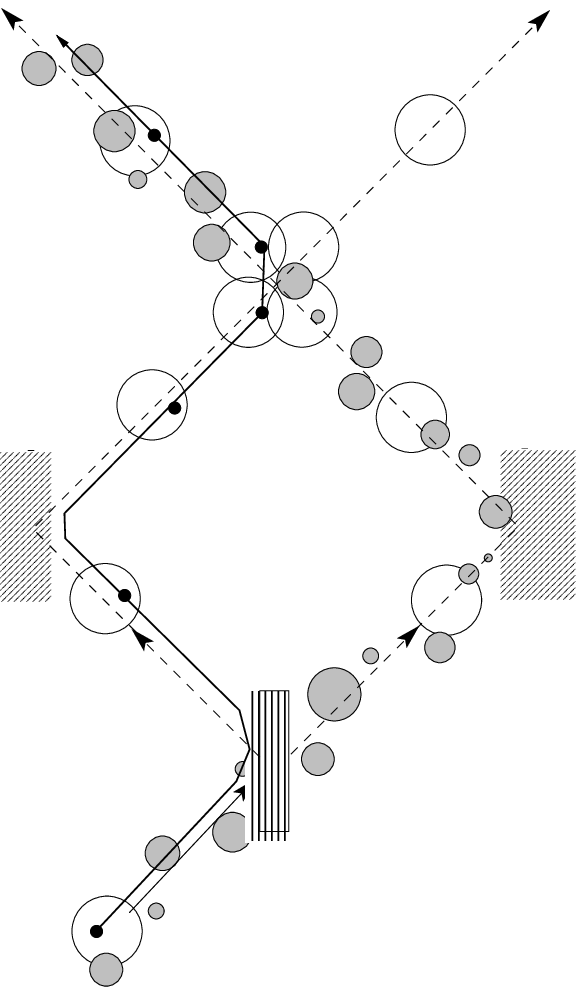} \end{center}

\noindent 
{\small Fig. 5. The trace of (slow developing) bubbles shows
  trajectory which is different from the trajectory of the Bohmian
  particle. (Again, these are the bubbles of the branch of the wave
  function corresponding to a particular world.) This is an example in
  which the ``world'' of the MWI is different form the ``world'' of
  Bohmian Mechanics: in the Bohmian world the particle moved in the left arm of the interferometer, while in the (postselected) MWI world it moved in the right arm.}

\vskip 1cm

\section{ Protective Measurements}

Another situation in which position measurements do not show Bohmian
positions are {\it weak measurements} \cite{AV90,AVBohm} and in
particular, weak adiabatic measurements of position of a particle in
nondegenerate energy eigenstates. Such measurements  are called {\it protective
  measurements} \cite{PM,PM1,PMBohm1,PMBohm2}. In protective
measurements we find, at the end, the wave function of the particle.
In many energy eigenstates the Bohmian particle does not move, so it
seems that the local values of the wave function obtained in the
experiment arise without the Bohmian particle being at the vicinity of
this location. However, it is not obvious that the measuring
interaction in the process of the protective measurement does not move
the Bohmian positions in such a way that the results of protective
measurements could be explained as the time average of the presence of the
Bohmian particle in a particular place. There have been an extensive
analysis of this question and it has been shown that it is not the
case, i.e. that the spacial profile of the wave function is obtained
without the Bohmian particle being present in most of the regions of
the non-vanishing wave function.

Consider a particle in a potential well, whose initial wave function
is the ground state, Fig. 6. We assume that Bohmian position is in point $A$ and we want to measure the density of the quantum wave at point $B$. If we introduce an adiabatic and weak
perturbation of the potential which eventually goes to zero, we
know that the wave function coincides at any moment with the ground
state of the instantaneous Hamiltonian (we assume that the ground
state is always nondegenerate). Our assumptions about the
perturbation  which is required for performing protective measurement of the
density of the wave function at the vicinity of $B$ ensure that the change in the wave function is small
at all times and eventually vanishes. The lemma proved by Aharonov et al. \cite{PMBohm2}  tells us that
the change in particle position is likewise small at all times. Thus, the perturbation of the potential at the vicinity of $B$ due  to the measurement will not change the Bohmian position significantly (which was originally at $A$) and will not bring it to $B$. 
 So, for a
Bohmian particle in a given position, we can probe the
wave function in most other positions without the particle ever
being present there.

\break

\begin{center} \leavevmode \epsfbox{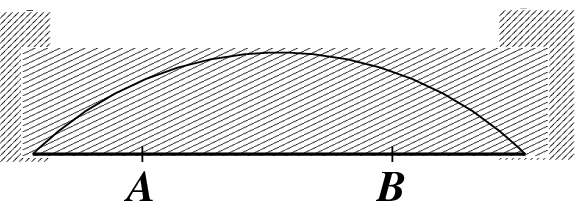} \end{center}

\noindent 
{\small Fig. 6. Ground state of a particle in a one-dimensional box.
  We can measure the density of particle's quantum wave at the
  vicinity of $B$ while the Bohmian position remains at the vicinity
  of $A$.}

\vskip 1cm

\section{Conditions for observable effects of empty waves}

To summarize above examples let us 
state clearly the conditions at which  empty waves cause an observable
effect. There are 
three conditions:

 i) Counterfactually, the wave should cause an
observable effect if at the particular time the Bohmian particle was inside it (i.e. the wave packet was a
non-empty wave). The meaning of ``observable effect'' is that some
other system changes significantly its quantum state.

 ii) At the time of the observation of the effect, the
Bohmian particle should be inside the wave. (At this later time the
wave is not in the interaction region, so the direct effect is
absent, and we still can consider it as an  effect  caused by an empty wave.) 

iii) The change of the quantum waves of other objects (the ``observable
effect'' of  (i)) should be such that  the spacial densities of their 
quantum waves  are not changed significantly: they  should not leave the
locations of the Bohmian positions of the undisturbed objects.

Trivially, (i) takes place in all our examples: in Surrealistic
trajectories spins are flipped, in the bubble chamber experiment the
bubbles leave  an observable trace and in the protective measurement,
the pointer of the measuring device changes its state. 

Figures 4 and 5 show that condition (ii) is satisfied both in the spin and in the bubble
chamber  experiments. In both cases, at the end of the experiment, the Bohmian particles inside the wave packet which was an empty wave packet before.

In the experiment with spins, the spacial wave function of the spin
particles remains without any change, i.e. (iii) is fulfilled. In the
bubble chamber experiment there is some change in the spacial wave
function of the particle, but it is insignificant. Indeed, (iii) is fulfilled 
 due to the condition of the  fast moving particle and slow
developing bubbles.

In the analysis of protective measurement, there is a difficulty with
defining ``empty'' and ``non-empty'' wave packets. We have to divide the
quantum wave of the particle into two parts: one includes point $A$, the
location of the Bohmian particle, and another includes point $B$ where
the measurement is performed. The problem is that, while the total
wave function is essentially constant, the wave packet which is  the part of the complete wave evolves in a
nontrivial way. In particular, whatever part including $B$ we take,
it will very soon evolve and reach $A$, i.e., it will cease to be an
empty wave. This explains how (ii) is fulfilled in protective
measurements. The basic property  of weak measurements is that  the
position of the pointer
of the measuring device has large quantum uncertainty (it is necessary
for having small value of the conjugate momentum which appears in the
interaction Hamiltonian). Thus, protective measurements fulfill
property (iii).

So, can an empty wave of a bullet kill? The answer is that only a very
special bullet can do this. First, it should later reach the location
of its Bohmian position. It sounds as a difficult, but not impossible
task. Second, it should not cause immediate change of Bohmian positions
of particles in the cat's body, i.e., until the bullet reaches its
Bohmian position. This tells us that the bullet cannot be a usual
bullet, which makes holes immediately after it passes through the
body. One might imagine that the bullet is just a single very fast
particle. But then property (i)  can hardly  be satisfied. A
single particle passing through a body does not kill.

\vskip 1cm

\section{Interpretation}

In Surrealistic Bohm trajectories \cite{Sur} as well as in the other examples
 described above, 
 a seemingly correct experiment shows one trajectory, while
 calculations yield that Bohmian trajectory is different.
   Nevertheless, I do not see a direct contradiction with
the  minimalistic approach to Bohmian theory in which our experience supervene solely on Bohmian position.  I believe that a Bohmian
 proponent has a good defense in the following argument: conceptually,
 in the framework of the Bohmian theory these experiments are {\it
   not} good verification measurements. Prediction of Bohmian theory
 for the motion of the particle is a vector function of time
 $\vec{r}(t)$. To test it we have to test the location of the particle
 at different times. Since ``reality'' corresponds only to Bohmian
 positions, we have to read the locations using Bohmian positions of
 the measuring device at that time. In all our surprising examples
 Bohmian particles of measuring devices moved only much later, not at
 the time in which   the  particle position was observed.
 When the Bohmian position of the measuring device was measured at the
 same time (as in the example presented in Fig. 2), no surprising
 behavior was observed. So, the Bohmian picture in which  our
 experience supervene on Bohmian positions is consistent. There are no experiments
 in which a ``good'' measurement of position (a measurement that records
 the position of a particle at a particular time using Bohmian
 positions of the measuring devices), shows  results which are inconsistent with
 calculated Bohmian positions.

 Still, these surprising examples make
 Bohmian approach less attractive: We see that there are important
 causal structures which cannot be explained using Bohmian positions
 alone, without explicit description of the quantum wave. For me it
 adds to the objections of the proponent of the MWI to the Bohmian
 approach. It leaves in the formalism the structure of all parallel
 worlds, but claims that they are not related to our experience. But
 in these empty worlds the wave in the shape of Lev Vaidman might also
 write a paper in the empty wave copy of the Foundations of Physics
 Journal, so how You, the reader, know that this is not such an empty
 wave world?

\vskip 1cm

\section{Appendix: Bedard's Arguments}

 The abstract
of Bedard's paper is:
\begin{quotation}
 According to the traditional presentation of Bohm's interpretation  we
have immediate epistemic access to particle properties but not
wave function properties, and mental states, pointer states, and ink
patterns supervene on particle properties alone. I argue that these claims
do not make physical sense, and I offer an alternative account that does.
\end{quotation}

What I accept or postulate (in the framework of my understanding of Bohmian mechanics) is that mental states supervene on particle
properties alone. My motivation is not to get ``classicality'' as
Bedard
suggests: the experiments show that Nature does not follow the laws of
classical physics, so there is no reason to put physics into a
classical picture. My reason to turn to Bohm is to find a way of
seeing a single world corresponding to the formalism of the physical
theory of the universe, since I see that many physicists and
philosophers have considerable difficulty with accepting numerous
parallel worlds which we do not observe directly.

Bedard's arguments have already been criticized by  Dickson
\cite{Dick}. As far as I can understand his philosophical jargon my refutation of Bedard is similar, but I believe it will be helpful to write here my arguments too.

Bedard's objection is that Bohmian positions at a single moment and
without additional information of the properties of the particles are
not enough to describe the reality. This is a correct statement, but
Bohmians do not claim the opposite.  The task of Bohmian (as well as
any other) interpretation is to find correspondence between the
mathematical formalism of the physical theory and our experience.
Since conscious experience requires some period of time, we {\it have}
to consider trajectories at some period of time and not just an instantaneous configuration
for describing (defining) objects. Thus, an object made out of
electrons only, in a configuration of a (real) cat made of electrons,
protons, neutrons, etc., will cease to have the configuration of a cat
long time before it can be perceived as a cat.  The configuration of
Bohmian particles have the shape of a cat for a considerable time if,
and only if, they related to the right kind of particles and they have
appropriate quantum wave. It is possible to imagine Universe with
different physical interactions in which my last statement is not
true. But for physical interaction we have in our Universe it is true.
Philosophical arguments of Dickson tell us that the situation in our
Universe is relevant.  Except for some very specific situations which
are difficult to arrange and which probably were never arranged in
real laboratories, everything we see or perceive in some other way is
described correctly by trajectories of Bohmian particles.

Bedard claims that there are problems also with color and television 
screen pictures. I do not think that it is so: I expect no
conceptual problem with defining Bohmian positions for photons.
However, I can also avoid this discussion using  the research of perception by our brains
made by Aicardi et al. \cite{GRWbrain} in order to answer the criticism of Albert and myself \cite{AlV} of a  Ghirardi, Rimini and Weber collapse proposal \cite{GRW}. We pointed out that in a Stern-Gerlach experiment in which
the particle with the spin hits a fluorescent screen, the GRW collapse
might not take place until the light from the screen comes to our eyes
in spite of the fact that macroscopic number of atoms become excited
in this process. Aicardi et al.\cite{GRWbrain} answered that inside the
brain, in the process of perception, numerous cells move macroscopic
distance depending on what we see, so at least inside the brain one
can find the shape of Bohmian particles corresponding to what we have
seen. Thus, it is feasible that  our mental
states  supervene on particle positions  alone.   

\vskip 1cm

\centerline{\bf  ACKNOWLEDGMENTS}
 
 It is a pleasure to thank Halina Abramowicz, Yakir Aharonov, Noam Erez and Shelly Goldstein for helpful discussions.  This research was supported in part
 by grant 62/01 of the Israel Science Foundation.


\vskip 1cm

\begin{thebibliography}{99}


\baselineskip 14pt

\bibitem{Bohm}
Bohm, D., Part I, Phys. Rev. \textbf{85}, 166; Part II, Phys. Rev. \textbf{85},180 (1952).

\bibitem{Gold} S. Goldstein, ``Bohmian Mechanics'', in {\it The Stanford
    Encyclopedia of Philosophy} (Winter 2002 Edition), Edward N. Zalta
    (ed.), URL =
    http://plato.stanford.edu/archives/win2002/entries/qm-bohm/.
  

\bibitem{Eve}
H. Everett, Rev. Mod. Phys. {\bf 29}, 454 (1957).

\bibitem{MWIStan}
 L. Vaidman, ``The Many-Worlds
Interpretation of Quantum Mechanics'', 
 {\it The Stanford Encyclopedia of
  Philosophy} (Summer 2002 Edition), Edward N. Zalta
(ed.), URL = http://plato.stanford.edu/entries/qm-manyworlds/.

\bibitem{AVBohm}
Y. Aharonov, and L. Vaidman,  in {\it Bohmian Mechanics and Quantum Theory:
An Appraisal}, J.T. Cushing, A. Fine, S.
Goldstein, eds. (Kluwer, Dordrecht  1996).

\bibitem{Bell}
J.S. Bell,      Int.  J. Quan. Chem. {\bf 14}, 155 (1980).
  
  \bibitem{Beda}
    K. Bedard,
    Phil.  Sci. {\bf 66}, 221 (1999).

\bibitem{Ho}
P.R.  Holland, {\it Quantum Theory of Motion}, (Cambridge: Cambridge University Press, 1993).


 \bibitem{Mad} 
T. Maudlin, {\it Quantum Nonlocality and Relativity}, (Oxford: Blackwell, 1994).


\bibitem{Alb}
D. Albert, {\it Quantum Mechanics and Experience} (Cambridge, MA: Harvard University Press, 1992).

  

\bibitem{Ha92}
L. Hardy,  Phys. Lett. A
{\bf 167},  11   (1992).


\bibitem{EV93}
A. C. Elitzur,  and L. Vaidman,   Found.  Phys.
{\bf 23}, 987 (1993).

\bibitem{Sur}
B.G. Englert, M.O. Scully, G. S\"{u}ssmann, and H.Z. Walther,
Naturforsch. {\bf 47a}, 1175 (1992).




\bibitem{PMBohm1}
Y. Aharonov, B.G.  Englert, and M.O. Scully,
 Phys. Lett. A
\textbf{263}, 137 (1999); \textbf{266}, 216 (2000).

\bibitem{PMBohm2}
Y. Aharonov, N. Erez, and M.O. Scully,
Phys. Scrip., to be published.

\bibitem{PM}
Y. Aharonov and L. Vaidman,  Phys. Lett. {\bf A 178}, 38
(1993).

\bibitem{PM1}
Y. Aharonov, J. Anandan,  and L. Vaidman,   Phys. Rev. {\bf A 47}, 4616 (1993).



\bibitem{ST1}
D. D\"urr, W. Fusseder, S. Goldstein, and  N. Zanghi, Z. Naturforsch. \textbf{48a}, 1261 (1993).

\bibitem{ST2}
C. Dewdney, L. Hardy, E.J. Squires, Phys. Lett. A \textbf{184}, 6 (1993).

\bibitem{ST3}
K. Berndl, M. Daumer, D. D\"urr, S. Goldstein, and N. Zanghi, Nuovo Cimento \textbf{110B}, 737 (1995).

\bibitem{ST4}
H. R. Brown, C. Dewdney, and G. Horton, Found. Phys. \textbf{25}, 329 (1995).

\bibitem{Bar}
J.A. Barrett
Phil. Sci. {\bf 67}, 680 (2000).


\bibitem{AV90}
Y. Aharonov,  and L. Vaidman,
 Phys. Rev. {\bf  A 41}, 11 (1990).



\bibitem{Dick}
    M. Dickson,
    Phil. Sci. {\bf  67}, 704 (2000).

\bibitem{GRWbrain}
F. Aicardi, A. Borsellino, G.C. Ghirardi, and R. Grassi,
Found. Phys. Lett. {\bf 4}, 109 (1991).

\bibitem{AlV}
D. Albert and L. Vaidman,
 Phys. Let. {\bf A 139}, 1 (1989).

\bibitem{GRW}
G.C. Ghirardi, A. Rimini, and T. Weber, 
Phys. Rev.{\bf  D
34}, 470 (1986).


\end{thebibliography}
\end{document}